\documentclass[sigconf]{acmart}
\usepackage{multirow}
\usepackage{adjustbox}
\usepackage{subcaption}
\usepackage{xcolor}
\usepackage{algpseudocode}
\usepackage{algorithm}

\AtBeginDocument{%
  \providecommand\BibTeX{{%
    \normalfont B\kern-0.5em{\scshape i\kern-0.25em b}\kern-0.8em\TeX}}}

\setcopyright{acmcopyright}
\copyrightyear{2018}
\acmYear{2018}
\acmDOI{10.1145/1122445.1122456}

\acmConference[Woodstock '18]{Woodstock '18: ACM Symposium on Neural
  Gaze Detection}{June 03--05, 2018}{Woodstock, NY}
\acmBooktitle{Woodstock '18: ACM Symposium on Neural Gaze Detection,
  June 03--05, 2018, Woodstock, NY}
\acmPrice{15.00}
\acmISBN{978-1-4503-XXXX-X/18/06}

\begin{document}



\title{Hysia: Serving DNN-Based Video-to-Retail Applications in Cloud}


\author{Huaizheng Zhang, Yuanming Li, Qiming Ai, Yong Luo, Yonggang Wen, Yichao Jin,}
\author{Nguyen Binh Duong Ta}
\affiliation{\institution{Nanyang Technological University, Indeed Inc., Singapore Management University}}
\email{{huaizhen001, yli056, qai001, yluo, ygwen}@ntu.edu.sg, jinyichao@indeed.com, donta@smu.edu.sg}


\begin{abstract}
Combining \underline{v}ideo streaming and online \underline{r}etailing (V2R) has been a growing trend recently. In this paper, we provide practitioners and researchers in multimedia with a cloud-based platform named Hysia for easy development and deployment of V2R applications. The system consists of: 1) a back-end infrastructure providing optimized V2R related services including data engine, model repository, model serving and content matching; and 2) an application layer which enables rapid V2R application prototyping. Hysia addresses industry and academic needs in large-scale multimedia by: 1) seamlessly integrating state-of-the-art libraries including NVIDIA video SDK, Facebook faiss, and gRPC; 2) efficiently utilizing GPU computation; and 3) allowing developers to bind new models easily to meet the rapidly changing deep learning (DL) techniques. On top of that, we implement an orchestrator for further optimizing DL model serving performance. Hysia has been released as an open source project on GitHub, and attracted considerable attention. We have published Hysia to DockerHub as an official image for seamless integration and deployment in current cloud environments.

\end{abstract}




\maketitle

\section{Introduction}

Recently, we have been witnessing the combined power of video streaming and e-commerce. Since online videos can reach millions of people, most companies have realized that they are the best showcase platforms to promote products. Therefore, many applications have been developed to support this combination. These applications includes fashion products retrieval from videos \cite{garcia2017dress}, contextual ads insertion \cite{chen2019livesense}, etc. We term them Video-to-Retail (V2R) applications. In V2R, the task is to analyze video content and products, and to match them to each other so that companies can promote products efficiently while maintaining the video watching experience of users.

Developing V2R applications is a non-trivial task. First, the data that will be fed into the application such as videos, product ads and product descriptions, is multi-modality. Processing, fusing and aligning these data to understand them better require much effort and are still very challenging \cite{baltruvsaitis2018multimodal}. Second, to match videos to products or vice versa in a non-intrusive way, accurate recognition and retrieval algorithms are needed. Third, processing speed is vital for maintaining a good user experience. A video usually contains hundreds of thousands of frames, and a product database may include thousands of items. How to efficiently process and match them remains an open problem.

To address these issues, representative works as listed in Table \ref{tab:v2o_compare} have considered the following two perspectives: the system perspective and the algorithm perspective. From the system perspective, for instance, Mei et al. \cite{mei2007videosense} build a system that includes a pipeline to unify the ads and video pre-processing for contextual ads insertion. In \cite{garcia2017dress}, they exploit the video frame redundancy and index features into a kd-tree for fast clothing retrieval. From the algorithm perspective, quite a number of matching frameworks employing DL models have been proposed. For instance, in \cite{cheng2017video2shop}, the authors design a framework consisting of an image feature network, a video feature network and a similarity network to match clothing in videos to online shopping images. In \cite{cheng2016video}, the authors use a set of models that include content understanding models to analyze user behavior, and video tags for accurate video advertising.

\begin{table*}[]
\centering
\caption{A comparison of Hysia and existing V2R related works.}
\label{tab:v2o_compare}
\begin{adjustbox}{width=1\textwidth}
\begin{tabular}{|l|c|c|c|c|c|c|c|c|}
\hline
\textbf{V2O related work} & \multicolumn{1}{l|}{\textbf{\begin{tabular}[c]{@{}l@{}}Product-to-\\ Video\end{tabular}}} & \multicolumn{1}{l|}{\textbf{\begin{tabular}[c]{@{}l@{}}Video-to-\\ Product\end{tabular}}} & \multicolumn{1}{l|}{\textbf{\begin{tabular}[c]{@{}l@{}}System\\ support\end{tabular}}} & \multicolumn{1}{l|}{\textbf{\begin{tabular}[c]{@{}l@{}}End-to-\\ end\end{tabular}}} & \multicolumn{1}{l|}{\textbf{\begin{tabular}[c]{@{}l@{}}Model\\ management\end{tabular}}} & \multicolumn{1}{l|}{\textbf{\begin{tabular}[c]{@{}l@{}}Model serving\\ optimization\end{tabular}}} & \multicolumn{1}{l|}{\textbf{\begin{tabular}[c]{@{}l@{}}Web interface\\ \&API\end{tabular}}} & \multicolumn{1}{l|}{\textbf{\begin{tabular}[c]{@{}l@{}}Open\\ source\end{tabular}}} \\ \hline
VideoSense. \cite{mei2007videosense}          & \checkmark                                                                                & $\times$                                                                                  & \checkmark                                                                             & \checkmark                                                                          & $\times$                                                                                 & $\times$                                                                                      & \checkmark                                                                                 & $\times$                                                                            \\ \hline
CAVVA. \cite{yadati2013cavva}               & \checkmark                                                                                & $\times$                                                                                  & \checkmark                                                                             & \checkmark                                                                          & $\times$                                                                                 & $\times$                                                                                      & \checkmark                                                                                 & $\times$                                                                            \\ \hline
Video eCommerce. \cite{cheng2016video}     & $\times$                                                                                  & \checkmark                                                                                & \checkmark                                                                             & \checkmark                                                                          & $\times$                                                                                 & $\times$                                                                                      & \checkmark                                                                                 & $\times$                                                                            \\ \hline
Garcia et al. \cite{garcia2017dress}        & $\times$                                                                                  & \checkmark                                                                                & \checkmark                                                                             & \checkmark                                                                          & $\times$                                                                                 & $\times$                                                                                      & \checkmark                                                                                 & \checkmark                                                                          \\ \hline
Video2shop. \cite{cheng2017video2shop}         & $\times$                                                                                  & \checkmark                                                                                & $\times$                                                                               & $\times$                                                                            & $\times$                                                                                 & $\times$                                                                                      & $\times$                                                                                   & $\times$                                                                            \\ \hline
Madhok et al. \cite{madhok2018semantic}        & \checkmark                                                                                & $\times$                                                                                  & $\times$                                                                               & $\times$                                                                            & $\times$                                                                                 & $\times$                                                                                      & $\times$                                                                                   & $\times$                                                                            \\ \hline
\textbf{Hysia (Ours)}              & \checkmark                                                                                & \checkmark                                                                                & \checkmark                                                                             & \checkmark                                                                          & \checkmark                                                                               & \checkmark                                                                                    & \checkmark                                                                                 & \checkmark                                                                          \\ \hline
\end{tabular}
\end{adjustbox}
\end{table*}


There is still much work to be done to make developing fast and efficient V2R apps in various domains easier. First, existing systems only focus on one kind of V2R application such as contextual video advertising (product-to-video) or retrieving products from videos (video-to-product); and neglect the similarities (i.e. data engineering, model processing and matching) between them. Thus, multimedia researchers have to go through all the infrastructure plumbing work and make duplicate efforts in the process. Second, current systems pay more attention to improving matching accuracy and do not address system optimization. Third, given that DL models are increasingly used to build V2R applications, how to deploy these models with ease has not been fully considered. Lastly, there has been no comprehensive open source V2R platform for non-experts with little machine learning (ML) knowledge, making it challenging for them to harness the power of AI.

To narrow these gaps, we develop Hysia, a fully open source and cloud-oriented framework that comprises widely used V2R models and optimized infrastructure services including data engine, model serving and content matching. It allows non-expert users to quickly make use of the built-in utilities to analyze V2R related data; and expert users to build or evaluate new, high performance V2R applications with ease. Essential features in V2R such as application management and new model binding are also provided. Hysia can run in either virtual machines (VMs) or containers, making it easy to be integrated into the current cloud environments. 

In Hysia, multimedia practitioners and researchers can focus on application design rather than writing repetitive codes, with reference applications provided out of the box. We integrate industry-grade libraries to speed up data processing including NVIDIA video SDK for video pre-processing, Facebook faiss for searching and gRPC for communication. Hysia is highly modular, allowing seamless integration with new modules. Though it has been designed for V2R, it can also be used as a multimedia toolbox for video analysis, audio recognition and so on.

We release Hysia as an open source project at \url{https://github.com/cap-ntu/Video-to-Retail-Platform} under Apache 2.0 license. It has attracted attention and interests from many in the developer community. We also dockerize the system and publish it to DockerHub at \url{https://hub.docker.com/r/hysia/hysia} so that any cloud users can install and run Hysia with ease.

\section{System Design}
In this section, we first present the architecture of Hysia, and then we introduce the workflow for fulfilling V2R applications. 

\subsection{Architecture}

\begin{figure}[!ht]
  \centering
  \includegraphics[width=\linewidth]{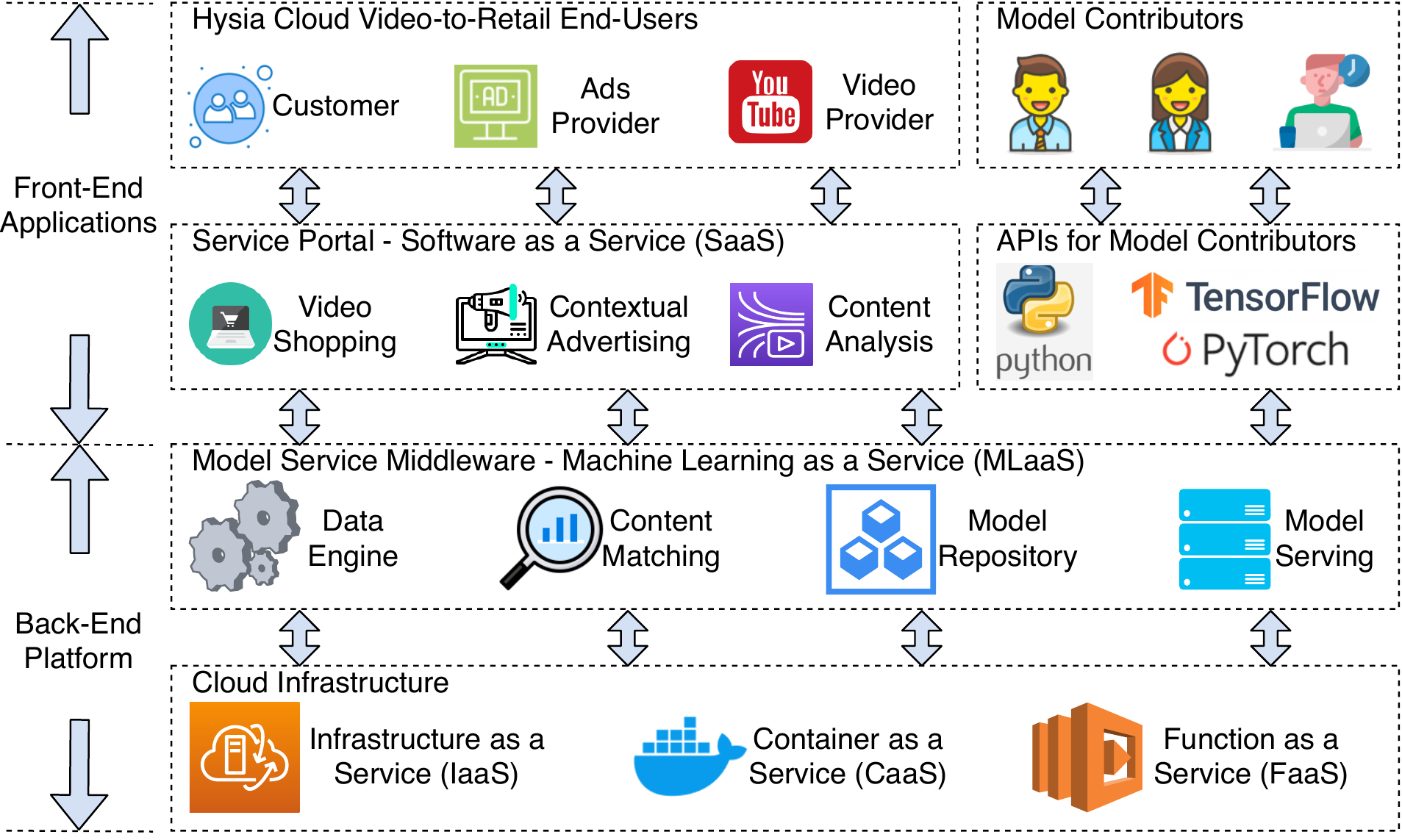}
  \caption{Hysia architecture.}
  \Description{The Hysia architecture.}
  \label{fig:v2o_arch}
\end{figure}

The system architecture is presented in Figure \ref{fig:v2o_arch}. In designing Hysia, we focus on the system's modularity and extensibility. As a cloud-oriented and end-to-end platform, it consists of two components: a back-end infrastructure, and a front-end application layer.

\textbf{Back-End Platform}. In clouds, computing resources are abstracted via three main approaches, namely infrastructure as a service (IaaS), container as a service (CaaS), and serverless/function as a service (FaaS). Hysia core services can make use of either virtual machines (IaaS) or containers (CaaS). In addition, as serving ML models is stateless \cite{zhang2019mark}, it is simple to deploy them using serverless (FaaS).

The ML model-related services, namely data engine, model repository, content matching and serving in Hysia are encapsulated into a middleware - a form of ML-as-a-Service. The data engine is designed to reduce users' efforts for preprocessing complex multimodality data. The model repository manages various ML models in Hysia. The model serving and content matching are designed to speed up the data analysis by utilizing GPUs. The functions provided by these core services are exposed via APIs so that developers can easily extend our system.

\textbf{Front-End Application}. Built on top of the back-end platform, the front-end application layer provides full support for four classes of users: 1) We have well-designed APIs for model contributors to bind new V2R-related models, develop new V2R applications and extend Hysia's functionalities; 2) We provide content analysis service for video providers so that they can mine videos to improve their commercial value; 3) A contextual advertising application is designed for advertisers to place ads at appropriate positions of videos; and 4) Hysia also has a video shopping service to help spectators buy products while watching videos. The built-in services and applications not only demonstrate the capability of our platform; they also provide reusable templates for researchers and practitioners to easily add more V2R plugins to Hysia to better serve their needs. 

\subsection{Workflow}

\begin{figure}[t]
  \centering
  \includegraphics[width=\linewidth]{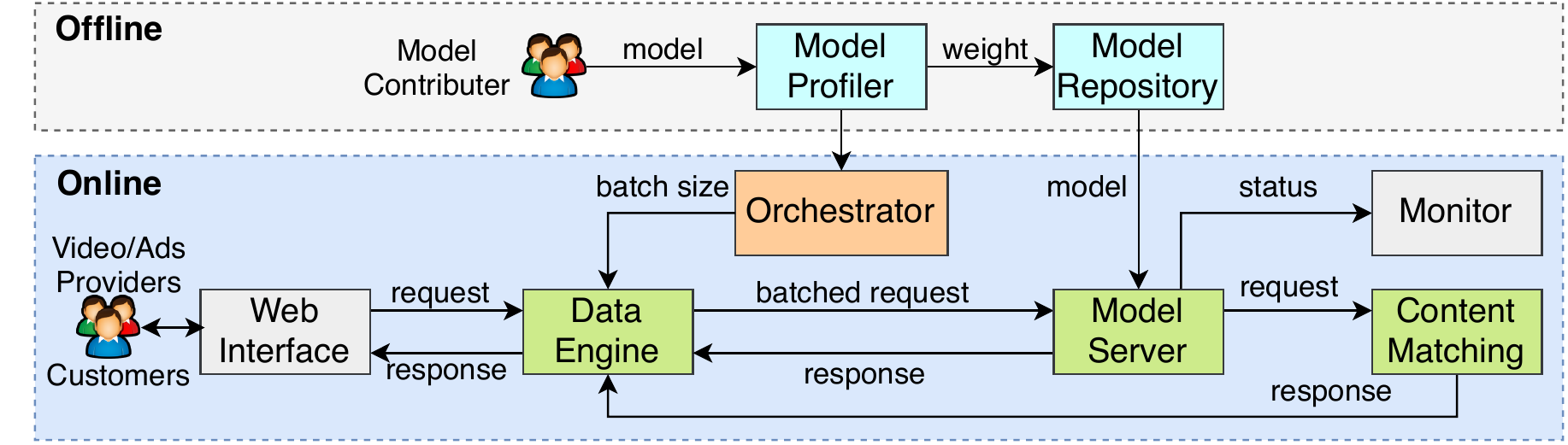}
  \caption{Workflow of a V2R application.}
  \Description{Hysia Workflow}
  \label{fig:v2o_workflow}
\end{figure}

The workflow of our system is illustrated in Figure \ref{fig:v2o_workflow}, which includes two phases: offline and online. In the offline phase, model contributors register their V2R-related models to Hysia and use the profiler to obtain their runtime performance. The profiling results are stored into a cache in the orchestrator; and the model weights are then persisted in the model repository.

In the online phase, a web interface is provided for end users (e.g., video providers and advertisers) to upload data (e.g. videos and ads), and to display final results. Those data are first preprocessed by the data engine and transformed into formats acceptable to DL models. Meanwhile, the orchestrator sends the optimal batch size of a model to the data engine so that it can batch the formatted requests. They are then fed into the model server for further analysis. Finally, the predictions and data feature output from the model server will be sent to the content matching service to do matching of videos to products or vice versa. We also implement a monitoring component to record the system status.

\section{System Implementation}
In this section, we describe the implementation of Hysia as illustrated in Figure \ref{fig:v2o_workflow}.

\textbf{Model Repository}. Hysia stores ML models in a two-layer structure. It persists model information such as the model name, the service description (e.g., product detection) and so on in SQLite which is a very lightweight database. The simple data structure makes it easy for users to replace the storage backend with their own database solutions. The model weight file, usually sizeable, is serialized and stored separately in a file system and the file path will be persisted in SQLite.

\textbf{Model Profiler}. This component receives ML models submitted by contributors and profiles these models offline. Much research has shown that the batch size can significantly impact the model's latency and throughput when served, in fact, our experiments also demonstrated that clearly in Section \ref{sec:expriment}. Therefore, Hysia profiles models under different batch sizes to obtain the corresponding latency and throughput. The profiling information will be stored in a cache in the orchestrator to help users choose the best batch size for a particular model.

\textbf{Orchestrator}. The orchestrator contains a cache implemented with Redis to store the model profiling information, and a batch size calculator for selecting an appropriate batch size of a model. Expert users of Hysia only need to specify the maximum acceptable latency for their applications, i.e., a latency SLO (Service-Level-Objective). The orchestrator can then decide on an appropriate batch size, and sends such value to the data engine.

\textbf{Data Engine}. The data engine implements a set of functions to pre-process multi-modality data, such as video, audio, product images, and textual content. \textit{(1) Video}: we employ NVIDIA video SDK to implement the HysiaDecode component to process videos with GPUs. In addition to utilizing GPU, HysiaDecode can also detect scene changes quickly by processing only one key frame in a scene shot. \textit{(2) Audio}: we separate audio from video and save it as a file that will be processed by suitable audio models. \textit{(3) Image}: we provide a resize and transform function to format original images so that they can be processed by existing Tensorflow or PyTorch models. \textit{(4) Text}: we implement a function to convert subtitles into ordinary text format, and a set of text preprocessing utilities such as tokenization so it can be fed into NLP models.

\textbf{Model Server}. The model server is implemented using gRPC which is widely used for building micro-services. It receives batched data from the data engine and employs models in the repository to analyze them. The model server will output two kinds of results. One is the prediction, and the other is the intermediate features. The predictions will be sent back to the data engine for displaying to users. The feature vectors will be stored in the file system, and at the same time, be sent to a subsequent module for matching. 


\textbf{Matching}. We implement this module to match products to videos or vice versa. Much optimization has been done in Hysia to improve the matching efficiency. Specifically, we employ faiss \cite{johnson2019billion}, and load features into GPUs. Therefore, the similarity comparison between features has been accelerated to meet real-time latency requirements. In addition, to make the system extensible, we provide APIs for experts to extend the module to accommodate their needs.

\textbf{Monitor}. The monitor is implemented with a pub-sub structure in Redis to support V2R applications running on a distributed infrastructure. It aggregates workers' status including CPU and GPU data, and resource usage of the executing models periodically. A master worker is set up to collect monitoring data from all worker nodes, making it easy for users to locate system issues.


\section{Demonstration}

Hysia incorporates a wide range of ML models, ranging from scene recognition and object detection to celebrity recognition and audio recognition, for building comprehensive V2R applications. In this section, we describe two built-in reference applications\footnote{\url{https://cap-ntu.github.io/hysia_mm_demo/}} including contextual advertising and video shopping, based on real-world scenarios. Then, we demonstrate how to bind new V2R models in Hysia. Finally, we present a quantitative evaluation of Hysia.

\begin{figure}
\begin{subfigure}{.5\columnwidth}
  \centering
  \includegraphics[width=1.0\linewidth]{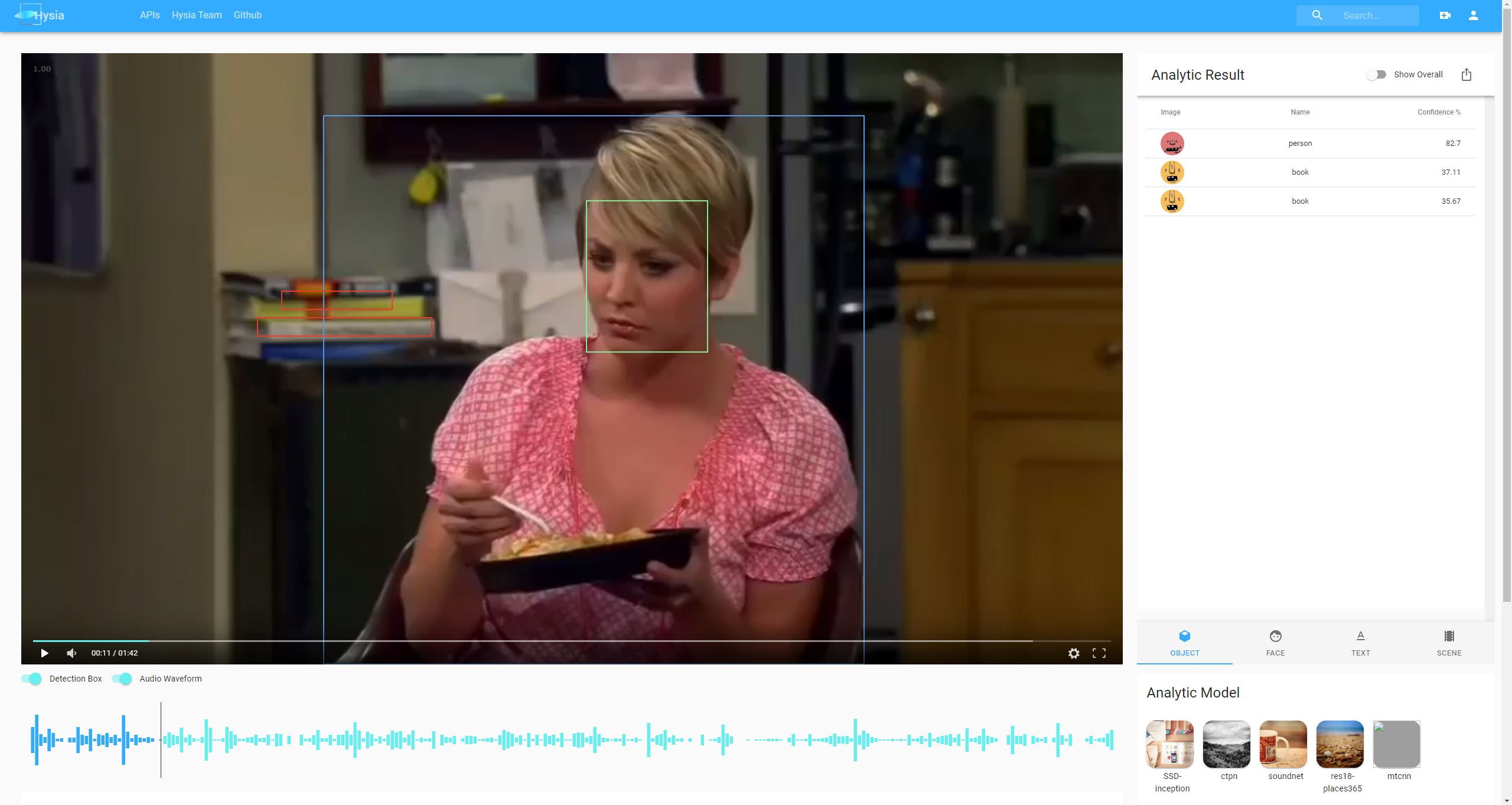}
  \caption{Video Analysis}
  \label{fig:video_analysis}
\end{subfigure}%
\begin{subfigure}{.48\columnwidth}
  \centering
  \includegraphics[width=1.0\linewidth]{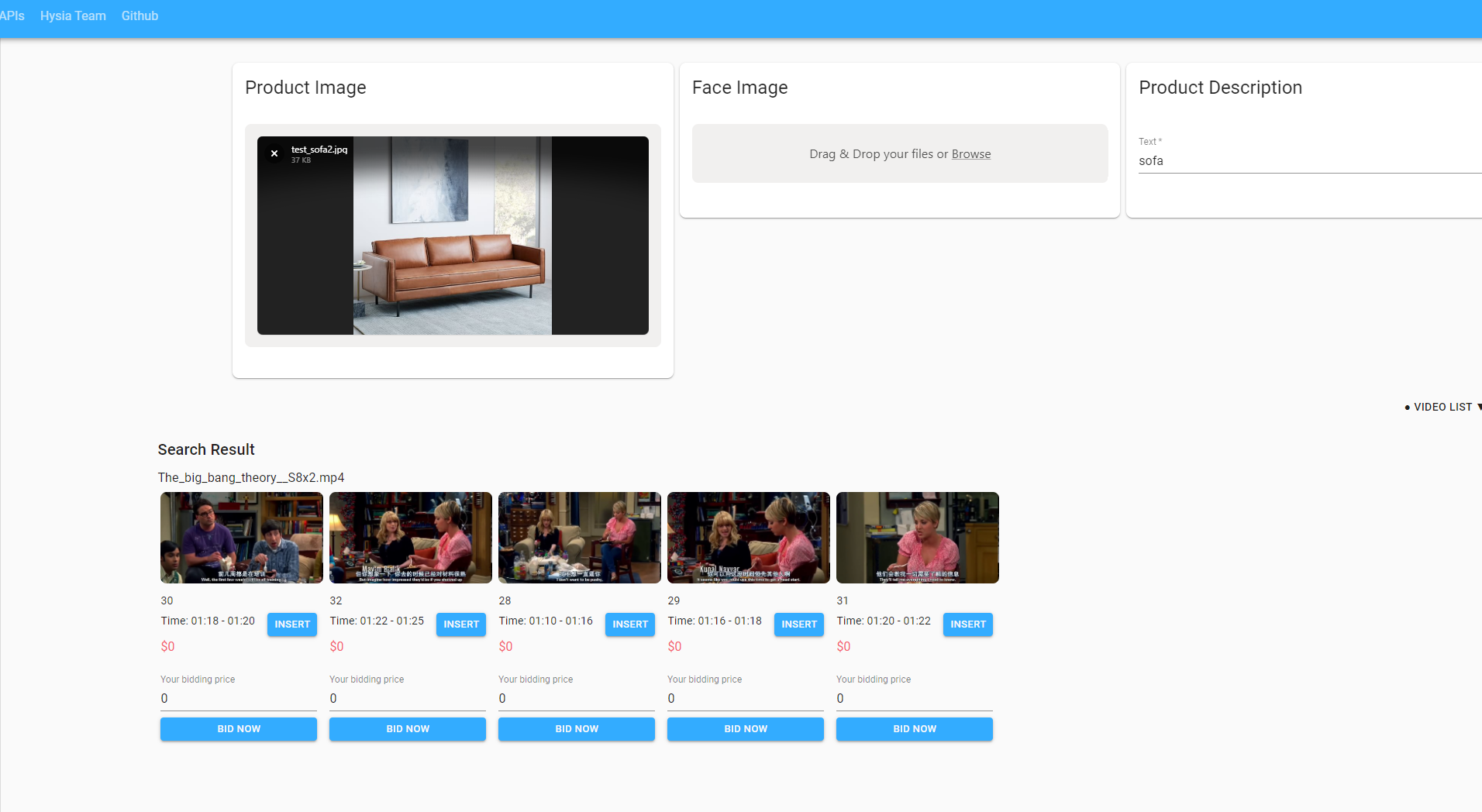}
  \caption{Ads Insertion}
  \label{fig:ads_insertion}
\end{subfigure}
\begin{subfigure}{.5\columnwidth}
  \centering
  \includegraphics[width=1.0\linewidth]{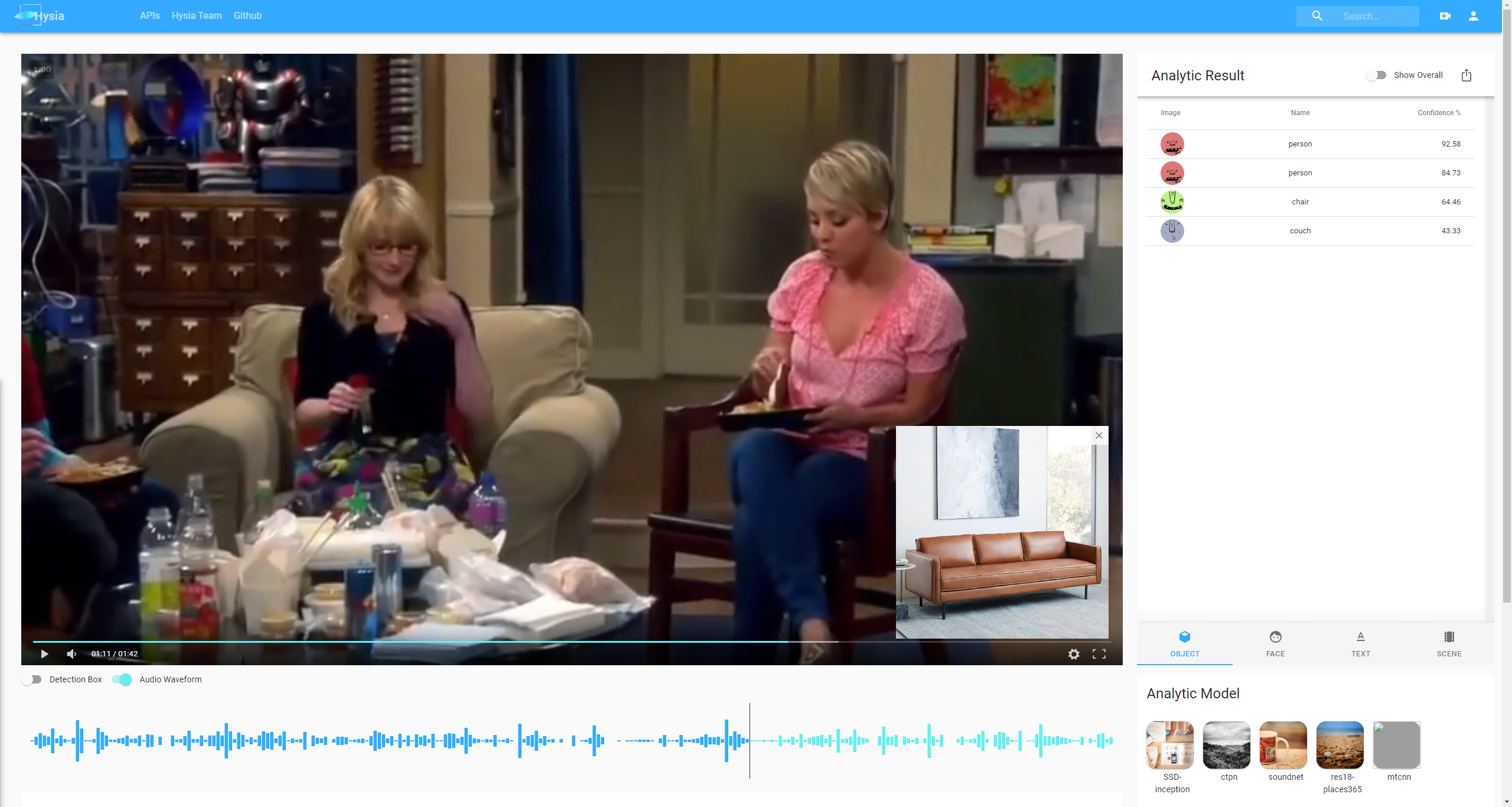}
  \caption{Ads Display}
  \label{fig:ads_display}
\end{subfigure}%
\begin{subfigure}{.47\columnwidth}
  \centering
  \includegraphics[width=1.0\linewidth]{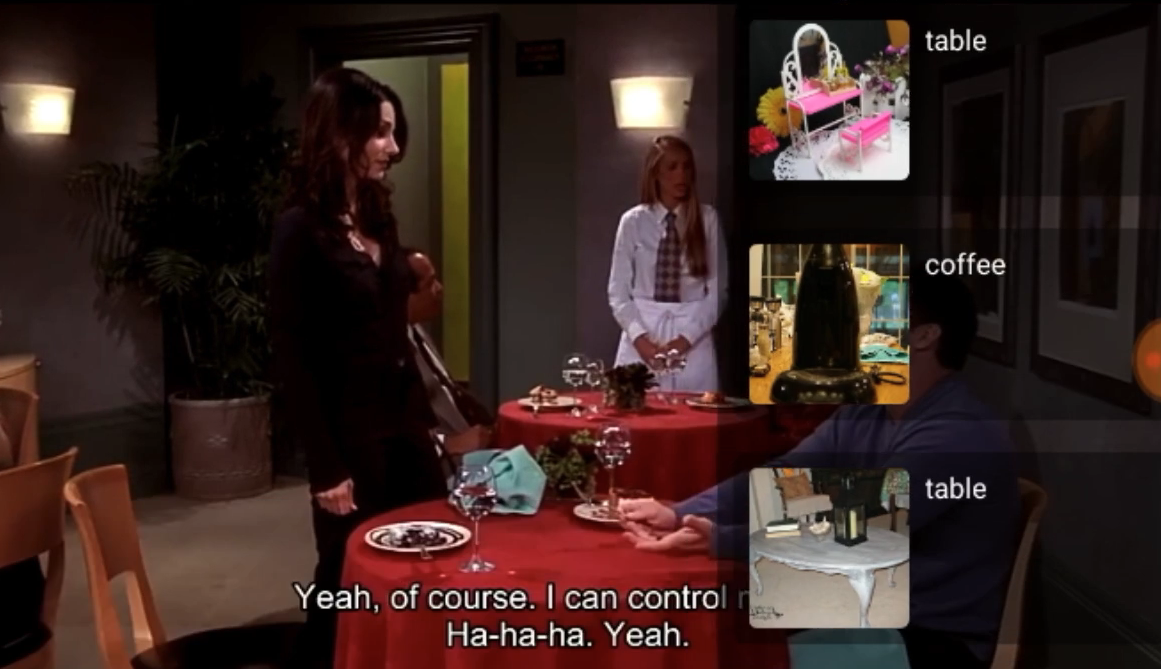}
  \caption{Video Shopping}
  \label{fig:video_shopping}
\end{subfigure}
\caption{The built-in applications of Hysia. Try them out yourself online.}
\label{fig:demo}
\end{figure}

\subsection{Contextual Advertising}

Both content and ads providers can enjoy the convenience provided by Hysia. For instance, a content provider gets a whole TV show and needs to insert several ad images or videos into the appropriate positions of videos. Hysia analyzes the uploaded video content as shown in Figure \ref{fig:video_analysis}. Then, advertisers can upload their ads to Hysia, and it will search for the top-5 of relevant video clips. Users can then choose the most relevant one (Figure \ref{fig:ads_insertion}). Here we leverage the human-in-the-loop factor, since real-world scenarios can be very complex. Automatically inserting into the top-1 clip may negatively affect users' experience, if the matching algorithm cannot capture new data distributions. Finally, Hysia allows both content and ads providers to verify the insertion results as shown in Figure \ref{fig:ads_display}.

\subsection{Video Shopping}

Spectators may choose to buy related products while watching videos. Hysia fulfills this need by providing a video shopping service. Since mobile video accounts for a significant portion of video traffic, we demonstrate a mobile application whose backend server is based on Hysia. As shown in Figure \ref{fig:video_shopping}, users can click on the screen, and Hysia will immediately search for products related to the scene that users are watching. The top 10 products will be shown to users. They can then click on the product icon to navigate to the corresponding shopping page.

\subsection{New Model Binding}

In Hysia, model contributors can use the provided APIs for binding new V2R models. Hysia provides well-designed template configuration files and reference models. For instance\footnote{\url{https://github.com/cap-ntu/hysia_mm_demo}}, a developer has trained a VQA model \cite{singh2018pythia} on a new V2R related dataset. The developer just needs to prepare a \texttt{YMAL} file and a \texttt{engine.py} file, following Hysia's template. The model will be containerized as a gRPC-based web service. Users can then employ the new model in Hysia to analyze V2R-related data.

\subsection{Quantitative Evaluation}
\label{sec:expriment}

In this section, we evaluate Hysia's performance\footnote{\url{https://github.com/cap-ntu/Video-to-Retail-Platform/tree/master/tests}} on the Stanford Online Product \cite{oh2016deep} and TVQA video \cite{lei2018tvqa} datasets with a DGX workstation with NVIDIA V100 GPUs.


\begin{figure}
\begin{subfigure}{.5\columnwidth}
  \centering
  \includegraphics[width=1.0\linewidth]{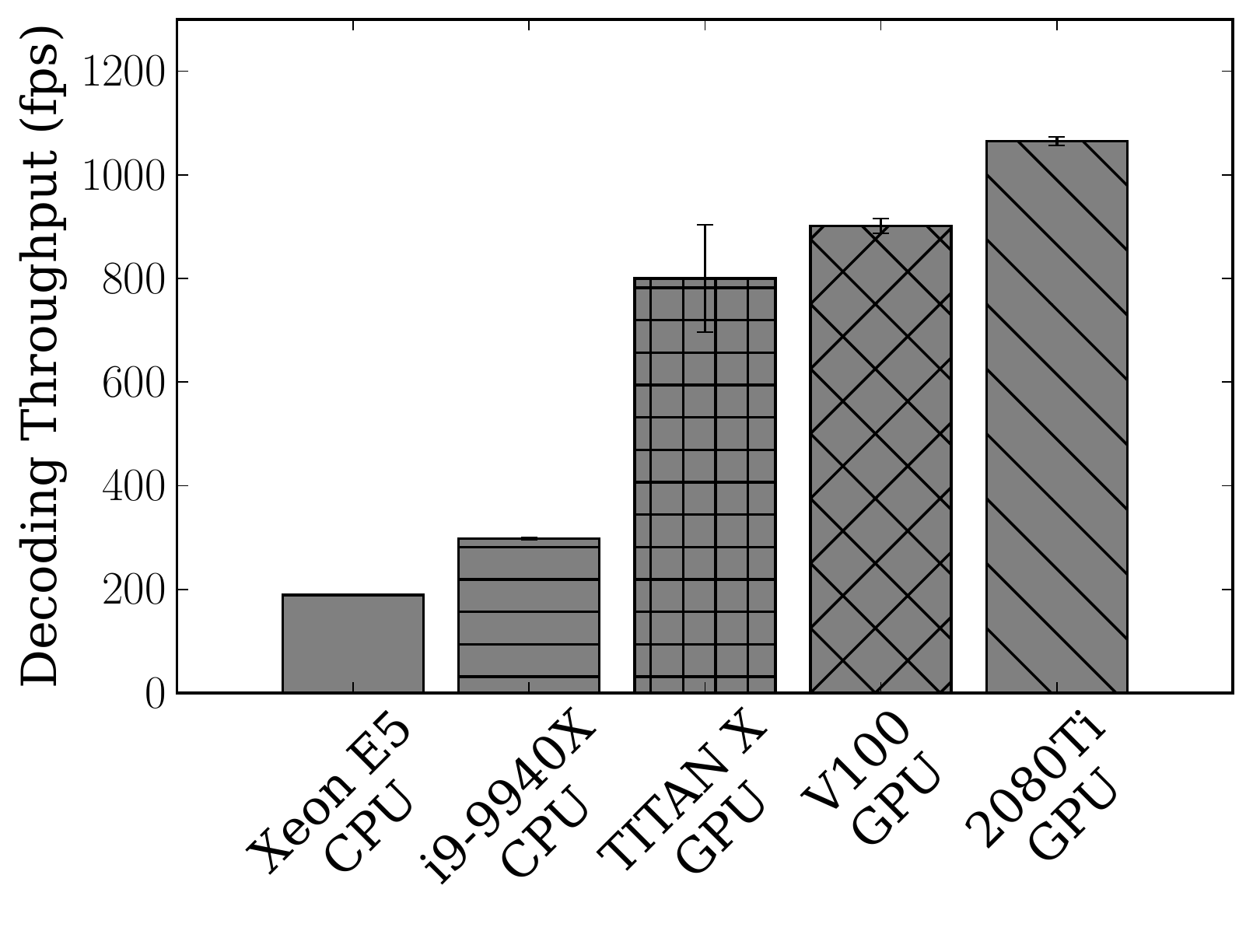}
  \caption{}
  \label{fig:video_throughput}
\end{subfigure}%
\begin{subfigure}{.5\columnwidth}
  \centering
  \includegraphics[width=1.0\linewidth]{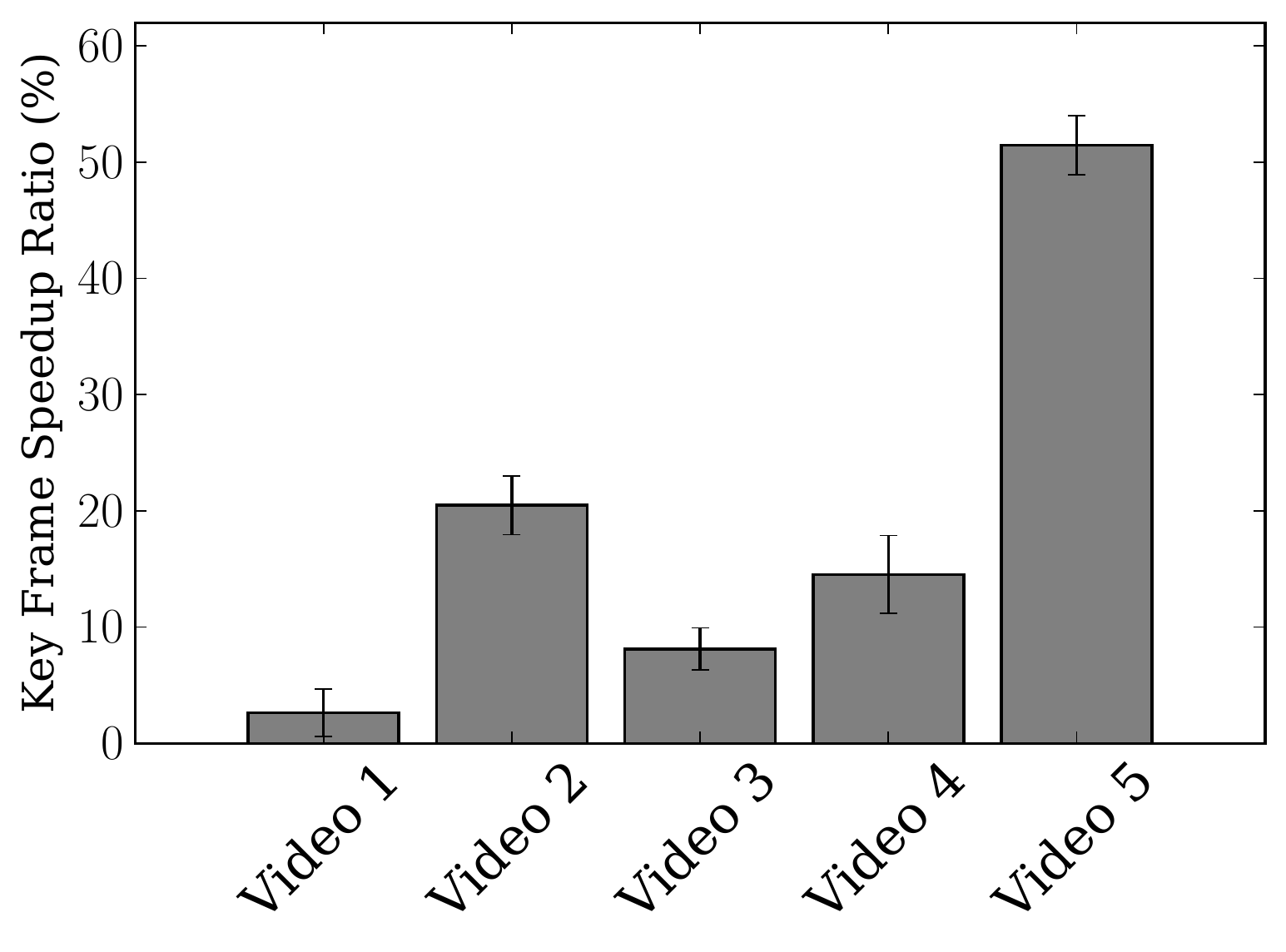}
  \caption{}
  \label{fig:match_latency}
\end{subfigure}
\begin{subfigure}{.55\columnwidth}
  \centering
  \includegraphics[width=1.0\linewidth]{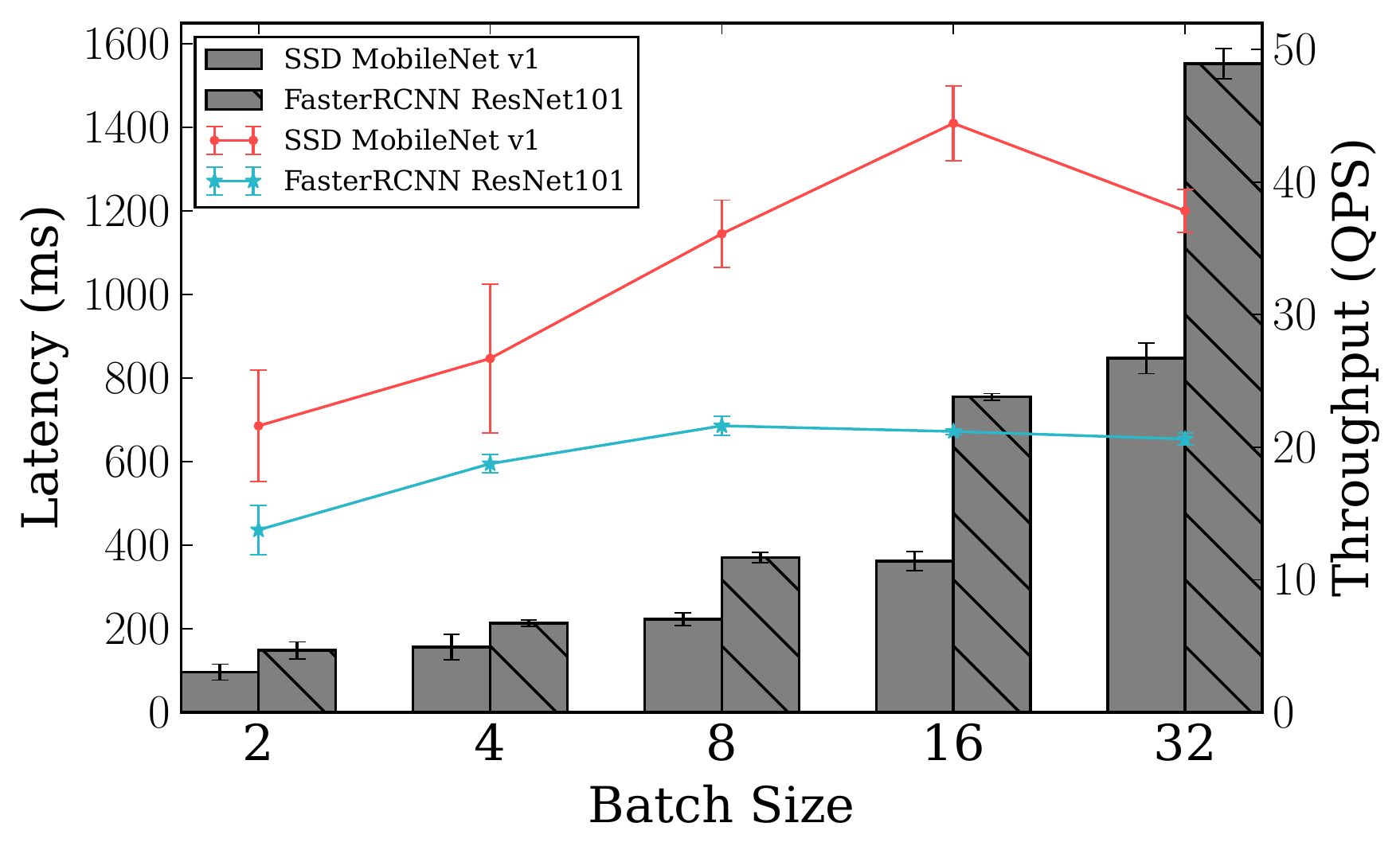}
  \caption{}
  \label{fig:model_throughput_latency}
\end{subfigure}%
\begin{subfigure}{.45\columnwidth}
  \centering
  \includegraphics[width=1.0\linewidth]{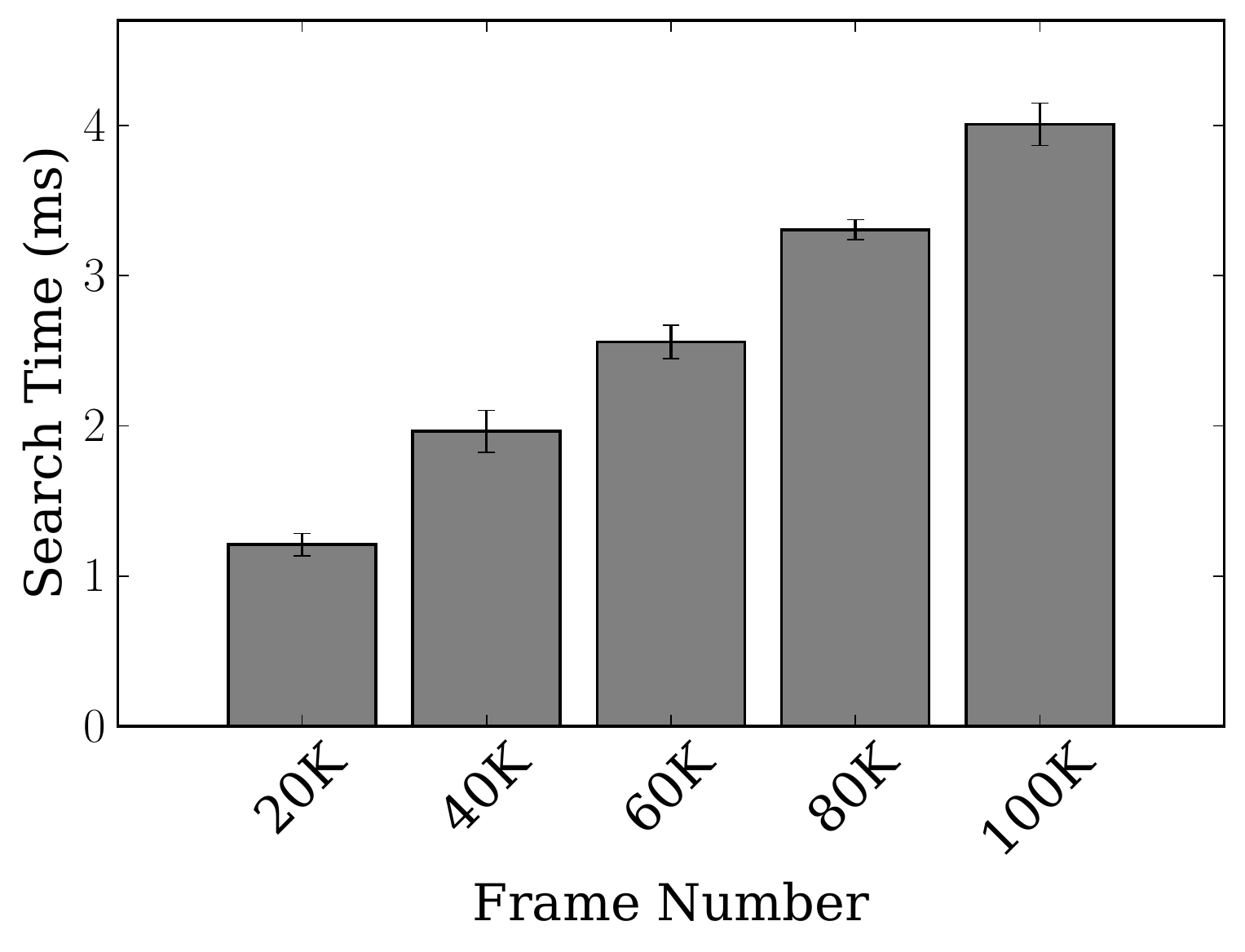}
  \caption{}
  \label{fig:memory_utilization}
\end{subfigure}
\caption{System performance evaluation}
\label{fig:system_evaluation}
\end{figure}

As shown in Figure \ref{fig:system_evaluation}, we have evaluated Hysia in four aspects: 1) Hysia data engine is able to efficiently utilize GPUs to process videos at the speed of more than 1000FPS, providing enough images for further analysis. 2) The key frame detection method can further improve video preprocessing speeds. A video with more scene shots can have more performance benefits. 3) As the batch size increases, the latency keeps increasing while the throughput increases initially, then decreases. This demonstrates the necessity of our model profiler and orchestrator for finding the right batch size. 4) By integrating faiss, Hysia's matching module can search for 100K of product images in less than 4.5ms. This demonstrates the ability to support a real-time shopping experience for spectators. 


\section{Conclusion}
\label{sec:conclusion}

In this paper, we present Hysia, a cloud-based system for the development and deployment of V2R applications. The system is designed to support a wide range of users, from ML novices to experts. The former can leverage built-in applications for V2R data analysis; while the latter can utilize Hysia's optimized services for rapid V2R prototyping. We demonstrate Hysia's usability with three real-world scenarios; and its efficiency with quantitative performance measurements. Our development team is continuously maintaining and improving Hysia as an open-source project.


\bibliographystyle{ACM-Reference-Format}
\bibliography{sample-base}

\end{document}